\documentclass[preprint,floats,aps,epsfig,nofootinbib,amssymb]{revtex4}
\usepackage{graphicx}
\usepackage{dcolumn}
\usepackage{bm}
\usepackage{mathrsfs}
\usepackage{slashed}
\usepackage{graphicx,color}
\usepackage{epsfig}
\usepackage{subfigure}
\usepackage{CJK}
\usepackage{xcolor}
\usepackage{amsmath}
\usepackage{amssymb}
\usepackage{caption2}
\usepackage{float}
\usepackage{amsmath}
\begin{document}

\title{Study of the rare decay $J/\psi\rightarrow e^+e^- \phi$ }
\author{Xing-Dao Guo$^1$}
\author{Si-Run Xue$^2$}
\author{Hong-Wei Ke$^3$}
\author{Xue-Qian Li$^1$}
\author{Qiang Zhao$^2$}
\affiliation{$^1$ School of Physics, Nankai University, Tianjin
300071, China \\
$^2$ Institute of High Energy Physics and Theoretical
       Physics Center for Science Facilities, Chinese Academy of Sciences,
       Beijing 100049, China\\
$^3$ School of Physics, Tianjin University, Tianjin, 300072, China}

\begin{abstract}

We study the decay process of $J/\psi\rightarrow e^+e^- \phi$ where the relatively clean electromagnetic (EM) transitions appear at leading order at tree
level while the hadronic contributions only emerge via hadronic loop transitions. We include the low-lying scalar $f_0(980)$ and pseudoscalar $\eta/\eta'$ as
the dominant contributions in the evaluation of the hadronic loop contributions.
It is found that the hadronic effects are negligible comparing with the EM contributions.
The decay width of $J/\psi\rightarrow e^+e^- \phi$ is determined to be about $2.12\times 10^{-6}$ keV if
there is no any other leading mechanism contributing, and this result will be
tested by the BESIII experiment with a large data sample of $J/\psi$.

\end{abstract}

\maketitle

\section{Introduction}

One of the advantages of high-intensity experiments is that by precision measurement one can testify the standard theory to high accuracy and then  look for
traces of new physics beyond the standard model (BSM). In the relatively low energy regime it is believed that the new physics effects may manifest themselves in rare decays
because the standard model contributions are highly suppressed in these channels, so that if the data exhibit an anomaly,
it would hint a possible contribution coming from the BSM physics. Even though such results may not pin down what kind of new physics plays a role, it
may offer valuable information about the BSM and complement the search for BSM phenomena at high energy frontier experiment such as LHC. For such a purpose
the BEPCII/BESIII~\cite{Ablikim:2012cn} and BELLE-II~\cite{Drutskoy:2006fg,Yuan:2007sj} experiments would provide large databases for $\psi$ and $\Upsilon$ families and enable us to investigate the rare processes which may expose traces of new physics BSM.

The process $J/\psi\rightarrow e^+e^-  \phi$ can be regarded as a rare decay process which can be measured by the BESIII Collaboration.
The leading order (LO)  contribution to this decay mode comes from the EM transition via $J/\psi\rightarrow e^+e^-\gamma^*\rightarrow e^+e^-  \phi$ where the conversion of $\gamma^*\rightarrow \phi$ is described by the vector meson dominance (VMD) transition~\cite{Guberina:1980dc}. The hadronic contributions become sub-leading in this process via the hadronic meson loops where the $c$ and $\bar c$ annihilation involves gluon exchanges and production of light flavored quarks at the quark-gluon level. This process should be complicated in terms of quark-gluon degrees of freedom. However, by knowing the radiative decays of $J/\psi \to \gamma P$ and $\gamma S$, $\phi \to \gamma P$ and $\phi\to \gamma S$ in experiment, where $P$ and $S$ denote the light-flavored pseudoscalar and scalar mesons, we can extract the relevant coupling vertices from experimental data and provide a reliable estimate of the hadronic contributions at hadronic level. We mention in advance that the hadronic contributions turn out to be smaller than the leading EM contribution and the most important intermediate hadronic contributions would come from the pseudoscalar or scalar mesons of which the masses are close to the final state $\phi$ meson. Notice that the intermediate states are produced by the hadronization of gluons. The isospin-1 states will be suppressed. Thus, the most possibly important intermediate states that are allowed would be pseudoscalar $\eta$ and $\eta'$ and scalar $\sigma(500)$ and $f_0(980)$. Taking into account the small coupling for $J/\psi\to \gamma \sigma(500)$ and $\phi\to \gamma\sigma(500)$ we can safely neglect the contribution from the intermediate $\sigma(500)$ in this leading order calculation.

The paper is organized as follows: in Sec. II and III, we derive the formulas for the contributions from
the LO EM process and hadronic meson loops, respectively.
In Sec. IV,  numerical results and analysis are presented. A brief summary will also be given there.

\section{The EM Contribution to $J/\psi\rightarrow e^+e^-  \phi$}

It is noted that the direct photon radiation of $J/\psi\rightarrow \gamma  \phi$ is strictly forbidden by the $C$-parity conservation
unless there is BSM new physics \cite{He} to break the rule.  In fact,
such a radiative decay has never been experimentally observed so far. But, when the photon is virtual
and later converts into an electron-positron pair, there is no restriction in principle and $J/\psi\to e^+e^-\phi$ can be measured.

\begin{figure}[H]
\centering
\begin{minipage}[!htbp]{0.6\textwidth}
\centering
\includegraphics[width=0.98\textwidth]{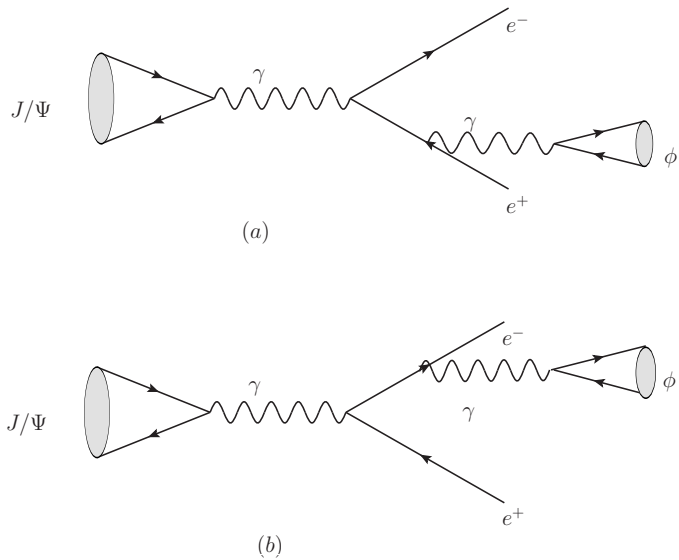}
\caption{The EM transitions of $J/\psi\to e^+e^-\phi$.}
\label{qed}
\end{minipage}
\end{figure}

We first compute the leading EM contribution to $J/\psi\rightarrow e^+e^-\phi$ for which the corresponding Feynman diagrams
are shown in Fig.~\ref{qed}.

In this process the couplings of $\gamma-J/\psi$ and $\gamma-\phi$ are
effective ones which we treat them at leading order in the framework of vector meson dominance (VMD).
Those effective vector currents can be expressed as the followings for the $\gamma-J/\psi$ and $\gamma-\phi$ couplings, respectively,
\begin{equation}
\begin{array}{rl}
&\langle J/\psi|\bar{c}\gamma^\nu c|0\rangle=g_{\psi\gamma}\varepsilon^{*\nu}_\psi\\
&\langle\phi|\bar{u}\gamma^\mu u|0\rangle=g_{\phi\gamma}\varepsilon^{*\mu}_\phi,\\
\end{array}
\end{equation}
where $\varepsilon^{\nu}_\psi$ and $\varepsilon^{\mu}_\phi$ are the polarization vectors of $J/\psi$
and $\phi$, respectively, and $g_{\psi\gamma}$ and $g_{\phi\gamma}$ are the corresponding couplings for $J/\psi$
and $\phi$ to a virtual photon, respectively, and they can be evaluated by data for $J/\psi$, and $\phi\to e^+e^-$~\cite{Agashe:2014kda}.

The Feynman amplitude corresponding to Fig.~\ref{qed} can then be obtained
\begin{eqnarray}
\mathcal{M}_{EM}&= &\bar{u}(p_1)[(-ie\gamma^\nu)\frac{\rlap /p_2+\rlap /p_\phi+m_e}{(p_2+p_\phi)^2-m^2_e}(-ie\gamma^\mu)\nonumber \\
&&+(-ie\gamma^\mu)\frac{-(\rlap /p_1+\rlap /p_\phi)+m_e}{(p_1+p_\phi)^2-m^2_e}(ie\gamma^\nu)]v(p_2)
\frac{1}{q^2_\phi}g_{\phi\gamma}\varepsilon^{\mu}_\phi\frac{1}{p^2_\psi}g_{\psi\gamma}\varepsilon^{*\nu}_\psi.
\end{eqnarray}

\section{The hadronic contribution to the Process $J/\psi\rightarrow e^+e^-  \phi$ via meson loops}

\begin{figure}[H]
\centering
\begin{minipage}[!htbp]{0.6\textwidth}
\centering
\includegraphics[width=0.98\textwidth]{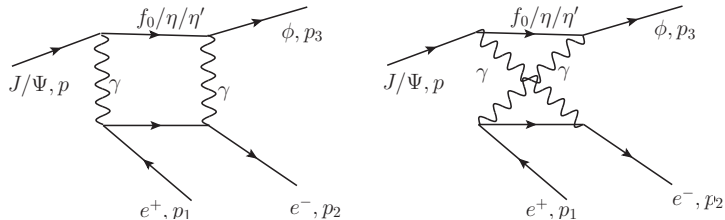}
\caption{Hadronic transitions contribute to $J/\psi\rightarrow e^+e^- \phi$ via meson loops.}
\label{qcd}
\end{minipage}
\end{figure}

The major hadronic contributions arise from the hadronic meson loops in $J/\psi\rightarrow e^+e^-  \phi$ as shown by Fig.~\ref{qcd}. Such non-perturbation effects can hardly be evaluated from the first principle. However, there are experimental data for $J/\psi \to \gamma f_0(980)$, $\gamma\eta$ and $\gamma\eta'$ as well as for $\phi\to \gamma f_0(980)$, $\gamma\eta$ and $\gamma\eta'$. We can then extract the vertex couplings from experimental data and estimate the leading hadronic contributions.


The effective couplings for vector meson radiative decays into scalar meson or pseudoscalar mesons are well established in the literature~\cite{Chung:1993da}.
The following couplings are adopted for which the Lorentz gauge invariance is automatically fulfilled:
\begin{eqnarray}
V_{J/\psi f_0\gamma} &=& g_{\psi f_0}\varepsilon_\psi\cdot\varepsilon_\gamma ,\nonumber\\
V_{\phi f_0\gamma} &=& g_{\phi f_0}\varepsilon_\phi\cdot\varepsilon_\gamma ,\nonumber\\
V_{J/\psi \eta\gamma}&=& g_{\psi \eta}
\varepsilon^{\lambda\mu\alpha\beta}p_\psi^\lambda \varepsilon_\psi^\mu (p_\psi-k_\gamma)^\alpha\varepsilon_\gamma^\beta ,\nonumber\\
V_{\phi\eta\gamma} &=& g_{\phi \eta}
\varepsilon^{\lambda\mu\alpha\beta}p_\phi^\lambda \varepsilon_\phi^\mu(p_\phi-k_\gamma)^\alpha\varepsilon_\gamma^\beta ,\nonumber\\
V_{J/\psi\eta'\gamma}&=& g_{\psi \eta'}
\varepsilon^{\lambda\mu\alpha\beta}p_\psi^\lambda \varepsilon_\psi^\mu (p_\psi-k_\gamma)^\alpha\varepsilon_\gamma^\beta ,\nonumber\\
V_{\phi\eta'\gamma} &=& g_{\phi \eta'}
\varepsilon^{\lambda\mu\alpha\beta}p_\phi^\lambda \varepsilon_\phi^\mu(p_\phi-k_\gamma)^\alpha\varepsilon_\gamma^\beta,
\end{eqnarray}
where $\varepsilon_\gamma$ is the polarization vector of the photon and the couplings can be extracted from either $V\to \gamma S$ or $V\to \gamma P$ with $S$ and $P$ denote the final state scalar or pseudoscalar meson, respectively. In principle, one should include all the intermediate scalar and pseudoscalar mesons and even higher spin states in the estimate of the hadronic contributions based on the quark-hadron duality argument~\cite{Bloom:1971ye}. However, one notices that the largest contributions are from the low-lying intermediate states with relatively large couplings. Thus, as an estimate of the leading hadronic contribution we only consider the low-lying scalar and pseudoscalar states in the meson loops. We also note that contributions from the intermediate $\eta_c$ are also possible which, however, is negligibly small due to the extremely small exclusive decay of $\eta_c\to \gamma\phi$. So it is a reasonable approximation that we only consider the scalar $f_0(980)$ and pseudoscalars $\eta$ and $\eta'$ as the leading contributions to the hadronic loops.

The Feynman amplitude induced by the scalar loop involving $f_0(980)$ is
\begin{eqnarray}
\mathcal{M}&=&\displaystyle\int\frac{d^4k}{(2\pi)^4}
\bar{u}(p_2)(-ie\gamma^\mu)\frac{\rlap /k-\rlap /p_1+m_e}{(k-p_1)^2-m^2_e}(-ie\gamma^\nu)\bar{v}(p_1)\nonumber\\
&&\times \frac{1}{k^2}\frac{1}{(p_\psi-k-p_\phi)^2}\frac{1}{(p_\psi-k)^2-m_{f_0}^2}
\varepsilon_\phi^\mu\varepsilon^{*^\nu}_\psi g_{\psi f_0} g_{\phi f_0},
\end{eqnarray}
where the coupling constants $g_{\psi f_0}$ and $g_{\phi f_0}$ are determined by the experimental data for $J/\psi\rightarrow \gamma f_0(980)$
and $\phi\rightarrow \gamma f_0(980)$~\cite{Agashe:2014kda}, respectively.

The Feynman amplitude induced by the loop involving $\eta$ is
\begin{eqnarray}
\mathcal{M}&=&\displaystyle\int\frac{d^4k}{(2\pi)^4}
\bar{u}(p_2)(-ie\gamma^\mu)\frac{\rlap /k-\rlap /p_1+m_e}{(k-p_1)^2-m^2_e}(-ie\gamma^\nu)\bar{v}(p_1)\nonumber\\
&&\times\frac{1}{k^2}\frac{1}{(p_\psi-k-p_\phi)^2}\frac{1}{(p_\psi-k)^2-m_\eta^2} \varepsilon^{\lambda\mu\alpha\beta}\varepsilon^\lambda_\phi (p_\psi-k)^\alpha2p_\phi^\beta
\varepsilon^{\delta\nu\sigma\rho}\varepsilon^{*\delta}_\psi p_\psi^\sigma2k^\rho g_{\psi \eta} g_{\phi \eta},
\end{eqnarray}
where the coupling constants $g_{\psi \eta}$ and $g_{\phi \eta}$ are determined by the experimental data for
$J/\psi\rightarrow \gamma\eta$ and $\phi\rightarrow \gamma\eta$~\cite{Agashe:2014kda}, respectively. For $\eta'$ the amplitude has the same expression as that for the $\eta$ loop.

\section{Numerical results and discussions}

In this section, we present the numerical results along with all the necessary inputs which
are adopted from the Particle Data Group~\cite{Agashe:2014kda}. The following mass values are adopted in the calculation, i.e.
$m_\psi=3.097$ GeV, $m_\phi=1.020$ GeV, $m_{f_0(980)}=0.980$ GeV, $m_\eta=0.548$ GeV, and $m_{\eta'}=0.958$ GeV.

For the vector meson and photon coupling in the VMD it can be extracted by the vector meson leptonic decay, i.e.
\begin{equation}
\Gamma(V\rightarrow e^+ e^-)=\frac{\alpha_e}{m_V^3}g_{V\gamma}^2 \ .
\end{equation}
With the measured widths $\Gamma(J/\psi\rightarrow e^+ e^-)=5.52$ keV and $\Gamma(\phi\rightarrow e^+ e^-)=1.26$ keV~\cite{Agashe:2014kda},
we obtain  $g_{\psi\gamma}=0.150$ GeV$^2$ and $g_{\phi\gamma}=0.013$ GeV$^2$.

The experimental data for $J/\psi\to \gamma f_0(980)$ are not available. However, we can estimate the
effective coupling constant $g_{\psi f_0}$ with the data for $J/\psi\to \omega f_0(980)$ and $\phi f_0(980)$ in the VMD. Therefore, the coupling can be expressed as
\begin{equation}
g_{\psi f_0}\equiv \sum_V \frac{g_{J/\psi f_0 V} g_{V\gamma}}{m_V^2} \ ,
\end{equation}
where $V=\omega, \ \phi$ as the vector meson fields contributing to $J/\psi\to \gamma f_0(980)$. The strong coupling $g_{J/\psi f_0 V}$ is estimated by
\begin{equation}
\Gamma(J/\psi\rightarrow V f_0)=\frac{|\textbf{q}|}{8\pi m_\psi^2}g_{\psi f_0 V}^2 \ ,
\end{equation}
where $\textbf{q}$ is the three-vector momentum of the final state mesons in the c.m. frame of $J/\psi$. Again, the coupling $g_{V\gamma}$ will be determined by $V\to e^+e^-$ from experiment. With the experimental data for $J/\psi\to \omega f_0(980)$ and $\phi f_0(980)$, i.e. $\Gamma(J/\psi\to \omega f_0(980))=0.0130$ keV and $\Gamma(J/\psi\to \phi f_0(980))=0.0297$ keV, we obtain $g_{\psi f_0 \omega}=1.57\times10^{-3}$ GeV and $g_{\psi f_0 \phi}=2.46\times10^{-3}$ GeV. With the measured width $\Gamma(\omega\rightarrow e^+ e^-)=4.37\times10^{-5}$ keV,
we obtain  $g_{\omega\gamma}=5.35\times10^{-5}$ GeV$^2$.In association with the experimental information for $\omega\to e^+e^-$, we obtain
$g_{\psi f_0}=3.21\times10^{-5}$ GeV.

For the $\phi f_0\gamma$ coupling, with the measured width
$\Gamma(\phi\rightarrow \gamma f_0(980))=1.37$ keV~\cite{Agashe:2014kda}, we determine $g_{\phi f_0}=3.03\times 10^{-2}$ GeV$^{-1}$.

For the couplings arising from the pseudoscalar meson loops
$g_{\psi \eta}$ and $g_{\psi \eta'}$ are obtained from the $J/\psi$ radiative decay
$\psi\rightarrow \gamma\eta$ and $\gamma \eta'$ as
\begin{equation}
\Gamma(J/\psi\rightarrow \gamma P)=\frac{|\textbf{k}_P|^3}{3\pi}g_{\psi P}^2,
\end{equation}
with the three-vector momentum $|\textbf{k}_P|=\frac{m_\psi^2-m_P^2}{2m_\psi}$ determined in the $J/\psi$-rest frame; $P$ stands for $\eta$ and $\eta'$.
With the measured value of $\Gamma(J/\psi\rightarrow \gamma\eta)=0.103$ keV
and $\Gamma(J/\psi\rightarrow \gamma\eta')=0.479$ keV~\cite{Agashe:2014kda}, we obtain  $g_{\psi \eta}=5.36\times 10^{-4}$ GeV$^{-1}$ and
$g_{\psi \eta'}=1.28\times 10^{-3}$ GeV$^{-1}$. Similarly, with $\Gamma(\phi\rightarrow \gamma\eta)=55.8$ keV and
$\Gamma(\phi\rightarrow \gamma\eta')=0.266$ keV~\cite{Agashe:2014kda} we obtain $g_{\phi \eta}=0.105$ GeV$^{-1}$ and
$g_{\phi \eta'}=0.108$GeV$^{-1}$.

\begin{table}[htbp]
\caption{The decay widths and branching ratios of $J/\psi\rightarrow e^+e^- \phi$ contributed from the EM and hadronic processes.
\label{decaywidth}}
\begin{center}
\begin{tabular}[c]{|c|c|c|c|c|c|}\hline
                                   &EM        & via $f_0(980)$     &via $\eta$ &via $\eta'$ & total  \\\hline
$\Gamma(J/\psi\rightarrow e^+e^- \phi)$(keV)&$2.12\times10^{-6}$ &$2.00\times10^{-13}$ &$7.2\times10^{-11}$ &$1.12\times10^{-9}$ &$2.12\times10^{-6}$\\\hline
B.R. & $2.28\times 10^{-8}$ & $2.16\times 10^{-15}$ & $7.75\times 10^{-13}$ & $1.20\times 10^{-11}$ & $2.28\times 10^{-8}$ \\\hline
\end{tabular}
\end{center}
\end{table}

The hadronic loops are calculated by LoopTools~\cite{Hahn:1998yk} where cut-offs for both ultra-violet (UV) and infra-red (IR) divergence are embedded. In Tab.~\ref{decaywidth} the calculated partial width for $J/\psi\to e^+e^- \phi$ is listed in comparison with the exclusive contributions from the leading EM process and the subleading hadronic loop transitions. Since the hadronic loop contributions are much smaller than the leading EM contribution, the sum of all the amplitude is saturated by the EM process. It is interesting to note that the results show that the decay of  $J/\psi\to e^+e^- \phi$ is highly suppressed at leading order and dominated by the EM contributions. In contrast, the hadronic contributions are even further suppressed by at least three orders of magnitude. It makes this process an ideal place where the SM background is rather small. Thus, it may serve as a potential process to probe effects due to BSM mechanisms. We point out such an advantage of this process but would not continue in this direction in this work to discuss expectations arising from BSM mechanism.


In summary, we investigate the decay mechanisms for $J/\psi\to e^+e^-\phi$ which appears to be a rare decay process in the SM. In particular, the leading contribution is from the EM transition at tree level of which the partial decay width is about $2.12\times10^{-6}$ keV. The subleading contributions from the hadronic meson  loops are found negligibly small and such an observation may make this decay process a useful probe for the search for BSM sources which can contribute to this process sizeably.

\section*{Acknowledgments}

This work is also supported, in part,
by the National Natural Science Foundation of China (Grant Nos.
11375128 and 11425525), and the Sino-German CRC 110 ``Symmetries and
the Emergence of Structure in QCD" (NSFC Grant No. 11261130311).

\end{document}